%% file: templateArxiv.tex
\documentclass{article}

\usepackage{PRIMEarxiv}
\usepackage{cite} 
\usepackage{multirow}
\usepackage{tabularx} 
\usepackage{subcaption}  
\usepackage{caption}  
\usepackage[utf8]{inputenc} 
\usepackage[T1]{fontenc}    
\usepackage{hyperref}       
\usepackage{url}            
\usepackage{booktabs}       
\usepackage{amsfonts}       
\usepackage{nicefrac}       
\usepackage{makecell}    
\usepackage{xcolor}
\usepackage{microtype}      
\usepackage{lipsum}
\usepackage{fancyhdr}       
\usepackage{graphicx}       
\graphicspath{{media/}}     
\usepackage{fontawesome5} 
\usepackage{amsmath}       

\makeatletter
\renewcommand{\thanks}[1]{%
  \protected@xdef\@thanks{\@thanks
    \protect\footnotetext{#1}}%
}
\makeatother

\title{PDD: Unleashing Economical and Flexible Heterogeneous LLM Inference via Cross-Datacenter Prefill-Decode Disaggregation
}

\author{
  Yida Wang$^{1}$, Xiuhong Li$^{1}$, Jianping Ma$^{1}$, Gan Sun$^{1}$, Yunshen Xu$^{1}$, Buhe Han$^{1}$, Jingxu Ng$^{1,4}$ \\ \textbf{Yuhao Luo}$^{1,4}$, \textbf{Ke Hong}$^{1,2}$,  
  \textbf{Guohao Dai}$^{1,3\text{\faEnvelope}}$, 
  \textbf{Boxun Li}$^{1 \text{\faEnvelope}}$, 
  \textbf{Yu Wang}$^{2 \text{\faEnvelope}}$\thanks{Corresponding to Boxun Li \texttt{liboxun@infini-ai.com}, Guohao Dai \texttt{guohao.dai@email.com}, and Yu Wang \texttt{yu.wang@email.com}} \\
  $^{1}$Infinigence-AI, $^{2}$Tsinghua University, $^{3}$Shanghai Jiao Tong University, $^{4}$Peking University
}

\begin{document}
\maketitle

\input{abstract}
\input{introduction}
\input{background}
\input{design}
\input{evaluation}
\input{discussion}

\input{conclusion}

\bibliographystyle{unsrt}  
\bibliography{references}

\end{document}

%% file: abstract.tex
\begin{abstract}

Interconnecting geographically dispersed clusters over wide‑area Ethernet provides a scalable and cost‑effective alternative to building dedicated intra‑datacenter heterogeneous clusters for Large Language Model (LLM) inference. Nevertheless, this disaggregated architecture across data centers imposes heavy KV‑cache communication between the two clusters, which in turn causes substantial Time‑to‑First‑Token latency. Such transmission latency is especially detrimental for agentic workloads, which are characterized by long contexts, high cache hit rates, and short output lengths.

To overcome these bottlenecks, we propose PDD on the basis of prefill-decode disaggregation.
PDD is a three-tier disaggregation architecture consisting of \underline{P}refill instance, Relay \underline{D}ecode (RLD) instance, and Main \underline{D}ecode (MD) instance.
RLD plays the important role to solve the KV cache communication latency issue.
On Cluster A, we deploy both Prefill instances and RLD instances. The Prefill instances handle the prefill‑phase computation, while the RLD instances receive the KV cache immediately via high‑speed RDMA after prefill completion. At the same time, the KV cache is also transmitted to Cluster B over Ethernet. This allows the RLD instances to begin decoding and overlap the KV transfer latency. On Cluster B, MD instances receive the KV cache from the Prefill instances over TCP‑based cross‑datacenter networks, along with the tokens produced by the RLD instances. The subsequent decoding tasks are then seamlessly handed off from the RLD instances to them for completion.

To maximize overall system efficiency, PDD employs three core mechanisms: Decode-side RadixCache to alleviate bandwidth bottlenecks, an Extend-Decode Handoff mechanism to ensure smooth control migration between RLD and MD, and multi-stage pipeline orchestration to manage complex dependencies between the 3-tiers under high concurrency and long-term serving. Finally, we design a low-cost, fine-grained heterogeneous deployment scheme that maximizes latency-masking efficiency at marginal cost, establishing a robust reference paradigm for future cross-datacenter inference systems. Compared to the Intra-DC homogeneous PD baseline, PDD's cross-datacenter mapping of compute-intensive H100s and memory-bandwidth-optimized H200s achieves a Benefit-Cost Ratio (BCR) up to 37.5\% higher in SLA-compliant goodput.
\end{abstract}

%% file: introduction.tex
\section{Introduction}
The rapid evolution of Large Language Models (LLMs) has driven an unprecedented demand for computational resources, making cost-efficient inference critically necessary\cite{Brown2020LanguageMA,Touvron2023Llama2O,Bai2023QwenTR,DeepSeekAI2025DeepSeekR1IR,Bai2026KimiKV,Zeng2026GLM5FV}. Since LLM inference consists of two computationally divergent phases—compute-bound Prefill and memory-bound Decode—modern accelerators exhibit highly asymmetric performance profiles across these phases. As our internal benchmarking demonstrates (detailed in Section~\ref{subsec:heterogeneous_pd}), specialized chips can achieve up to $2.1\times$ the throughput of an $8\times$ NVIDIA H100 baseline in Decode, yet lag significantly in Prefill. This inherent architectural disparity makes leveraging heterogeneous hardware a highly promising avenue, motivating heterogeneous Prefill-Decode (PD) disaggregation \cite{Patel2023SplitwiseEG,Zhong2024DistServeDP} to dedicate specific hardware types to the phases they optimize best. Despite these advantages, constructing dedicated heterogeneous clusters within a single datacenter incurs prohibitive capital expenditures and suffers from poor scalability. This inflexibility makes it difficult to adapt to the seamless shifting of model architectures and hardware landscapes. Pragmatically, an alternative approach is to interconnect geographically dispersed, already-deployed homogeneous clusters via low-cost wide-area Ethernet, enabling cross-datacenter heterogeneous inference.

However, this cross‑datacenter paradigm introduces severe communication bottlenecks, especially during Key‑Value (KV) cache transfer \cite{Qin2024MooncakeAK} in PD‑disaggregated inference. Compared with intra‑datacenter (intra‑DC) RDMA \cite{nvidia_gpudirect_rdma} networks, wide‑area Ethernet suffers from tens to hundreds of times higher transmission latency and substantially lower bandwidth. These limitations are particularly detrimental for agentic workloads \cite{Wang2023ASO}—a critical class of LLM applications characterized by extremely long contexts, high prefix cache hit rates, and short output lengths—where multi‑second to tens‑of‑seconds transfer delays become entirely unacceptable.

\begin{figure}[t]
\centering
\begin{subfigure}[t]{0.422\textwidth}
\centering
\includegraphics[width=\linewidth]{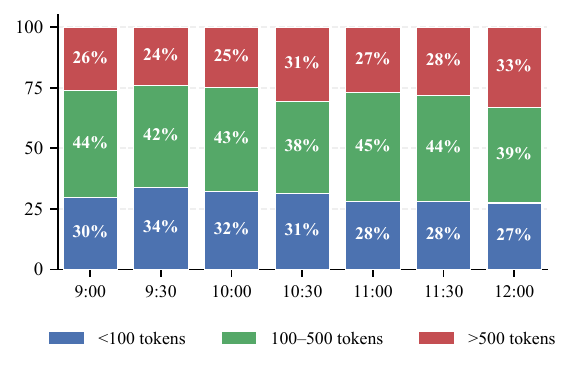}
\caption{Request output-length distribution.}
\label{fig:latency-constituents-a}
\end{subfigure}
\hfill
\begin{subfigure}[t]{0.528\textwidth}
\centering
\includegraphics[width=\linewidth]{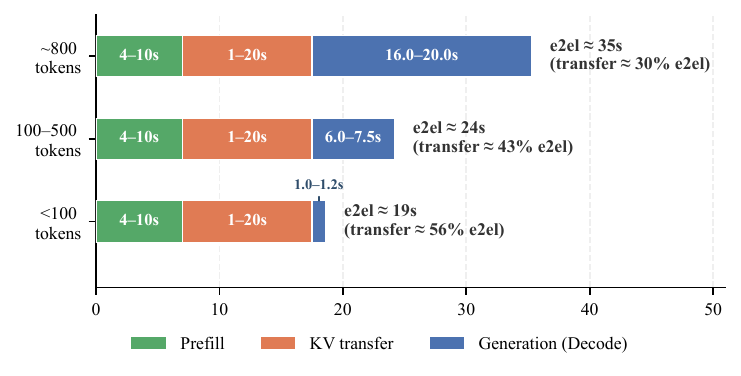}
\caption{E2EL constituents: prefill, KV transfer, and token generation.}
\label{fig:latency-constituents-b}
\end{subfigure}
\caption{%
(a)~In real-world agentic workload, 30\% of requests produce fewer than 100 tokens,
and 41.7\% produce 100--500 tokens. These short-output requests account for
over 71\% of total volume and remain stable across the 3-hour window encompassing 6M requests.
(b)~For these requests, the observed $1\sim 20\,s$ cross-datacenter KV transfer latency dominates the end-to-end latency. At 40$\sim$50 output
tokens per second, generation takes only $1\sim7.5\,s$; thus KV cache
transfer alone accounts for 43\%$\sim$56\% of e2el, making it the primary
bottleneck.%
}
\label{fig:latency-constituents}
\end{figure}

Specifically, the high cache hit rate significantly boosts the throughput of prefill instances, creating a severe supply-demand mismatch in bandwidth. For instance, a single cluster can process over 80 requests per second, with an average 64K context length. Even for advanced models like DeepSeek-V4-pro \cite{DeepSeekAI2026DeepSeekV4TH}, which employ techniques such as CSA and HCA to substantially compress the KV cache size, the total KV cache generated by 80 concurrent 64K-context requests still requires a minimum of approximately 23 GB of storage. Consequently, transferring this payload to the decode cluster demands a bandwidth of over 180 Gbps. In contrast, cost-effective cross-datacenter Ethernet connections typically provide bandwidths ranging merely from 10 Gbps to 100 Gbps. This substantial disparity establishes a critical bandwidth bottleneck for cross-datacenter heterogeneous inference. Beyond this bandwidth issue, cross-datacenter communication could also introduce unacceptable latency overheads. As depicted in Figure~\ref{fig:latency-constituents}, Ethernet-based KV cache transfer becomes the major constituent of the end-to-end latency (e2el) experienced by agentic applications, establishing itself as the primary performance bottleneck.

Despite these challenges agentic workload imposes, preliminary experiments (detailed in Section~\ref{subsec:non_uniform_latency}) show that cross-datacenter heterogeneous inference remains feasible. First, high prefix hit rates not only boost prefill instance throughput but also enable the Decode-side RadixCache (DRC) \cite{Zheng2023SGLangEE}. Through caching the historical KV cache on decode instances, DRC can reduce required cross-datacenter bandwidth by up to 10$\times$ at a 90\% average hit rate, effectively mitigating the bandwidth bottleneck. Second, as the Ethernet-based KV transfer micro-benchmarks show (see Table~\ref{tab:latency_distribution}), we observe that the severe KV transfer latency is non-uniform in practice. DRC minimizes payloads for high-hit-rate requests, which exploit the fairness mechanism of TCP to bypass congestion and arrive within $\sim$1 second. Extreme 20-second latencies are isolated to low-hit-rate requests with large payloads. Therefore, rather than addressing transfer latency globally, applying overlap techniques exclusively to these outliers provides a feasible solution.

Based on these insights, we propose \textbf{PDD}, a novel three-tier cross-datacenter architecture designed to enable economical and flexible heterogeneous inference. PDD establishes three distinct instances: Prefill (P) and RelayDecode (RLD) instances are co-located in the same datacenter, interconnected via a high-speed RDMA network; MainDecode (MD) instances, however, are situated in a geographically distant datacenter, connected to P and RLD via low-cost wide-area Ethernet (typically less than 20Gbps).

The PDD workflow operates as follows: Upon completing the prefill phase, the P instance simultaneously transfers the KV cache to an RLD instance via RDMA (taking tens of milliseconds) and to an MD instance via TCP over wide-area Ethernet (taking seconds). The RLD instance immediately initiates decoding upon receiving the KV cache, effectively masking the much slower Ethernet-based transfer. Once the MD instance completes the KV cache reception, it employs an \textbf{Extend-Decode Handoff} mechanism (detailed in Section~\ref{subsec:latency_masking}): it receives only the Token IDs generated by the RLD instance, recomputes their corresponding incremental KV cache locally, and seamlessly resumes token generation. This design enables PDD to achieve effective latency masking with minimal overhead. First, RLD instances consume marginal hardware resources, as they only generate partial output tokens for low-hit-rate requests, requiring far fewer resources than MD instances which handle the vast majority of token generation. Second, transmitting CPU-resident Token IDs and recalculating the incremental KV cache on memory-bound MD instances incurs negligible overhead. Crucially, this local recomputation approach is significantly more efficient than transmitting the incremental KV cache over Ethernet, delivering substantially lower and far more reliable latency.

Beyond the core architectural design of PDD’s three-tier disaggregation, we introduce several supporting techniques to enhance robustness, scalability, and cost-efficiency. To manage complex interactions and maintain DRC consistency, we design a multi-stage pipeline orchestration method. To further explore scalability, we leverage a wide-area Ethernet-based interconnect topology that enables flexible cross-datacenter routing (such as many-to-many topologies), allowing for horizontal scaling across geographically dispersed clusters. Additionally, we introduce a fine-grained heterogeneous deployment strategy—combining compute-heavy chips for P instances with a minimal fraction of memory-access-efficient chips for RLD instances within the same RDMA network—to mask cross-datacenter latency at a minimal cost while maximizing hardware utilization.

Extensive evaluations on DeepSeek-V4-pro using real-world agentic workload traces demonstrate that PDD effectively masks transmission latency and achieves superior cost-efficiency. Compared to the traditional Cross-Datacenter PD baseline, PDD reduces the P90 Time-To-First-Token (TTFT) by approximately 46\%. Microbenchmarks confirm that our local recomputation handoff eliminates bulky KV transfers and H2D/D2H overheads, capping maximum handoff latency at 321.4ms with minimal variance, compared to the 512.6ms peaks seen in naive Ethernet transfers. Furthermore, compared to the Intra-DC homogeneous PD baseline, PDD's cross-datacenter mapping of compute-intensive H100s and memory-bandwidth-optimized H200s achieves a Benefit-Cost Ratio (BCR) up to 37.5\% higher in SLA-compliant goodput.

%% file: background.tex
\section{Background}

\subsection{Motivating Heterogeneous Prefill-Decode Disaggregation}
\label{subsec:heterogeneous_pd}

LLM inference consists of two computationally divergent phases: Prefill, which is compute-bound, and Decode, which is memory-bound. As the AI hardware landscape rapidly diversifies, modern accelerators exhibit highly asymmetric performance profiles across these two phases. This inherent architectural disparity strongly motivates heterogeneous PD disaggregation, wherein specific hardware types are dedicated to the inference phases they are most optimized for.

To substantiate this motivation, we conducted comprehensive internal benchmarking across various accelerator architectures from multiple vendors. Evaluations were performed under a standardized agentic workload scenario: 64K input tokens, a 90\% prefix cache hit rate, and 512 output tokens, all while strictly adhering to Service Level Agreement (SLA) requirements. We measured the throughput of individual accelerator systems relative to a single $8\times$ NVIDIA H100 system baseline. The performance characteristics of these diverse chips are summarized in Table~\ref{tab:heterogeneous_bench}.

As Table~\ref{tab:heterogeneous_bench} illustrates, performance gaps among accelerators are substantial and highly variable. Vendor A (Chip A) achieves 56.7\% of the baseline's Prefill throughput and 38.7\% for Decode using the GLM5 model. Vendor C (Chip C), evaluated with GLM4.7, exhibits a different balance, offering 41.7\% for Prefill but a relatively stronger 56.4\% for Decode. Conversely, some architectures—such as those from Vendor B and Vendor D—often suffer from severe phase-specific bottlenecks, leading to significantly lower overall efficiency under homogeneous PD deployment.

\begin{table}[htbp]
  \centering
  \caption{Throughput of Diverse Accelerators Relative to an $8\times$ H100 Baseline (64K Input, 90\% Cache Hit, 512 Output)}
  \label{tab:heterogeneous_bench}
  \begin{tabular*}{0.95\textwidth}{@{\extracolsep{\fill}}lccc@{}}
    \toprule
    \textbf{Hardware}  & \textbf{Model} & \textbf{Prefill (\%)} & \textbf{Decode (\%)} \\
    \midrule
    Vendor A (Chip A)  & GLM-5   & 56.70\%  & 38.70\%  \\
    Vendor B (Chip B)  & GLM-5   & 43.50\%  & 15.50\%  \\
    Vendor C (Chip C)  & GLM-4.7 & 41.72\%  & 56.40\%  \\
    Vendor D (Chip D)  & GLM-4.7 & 24.70\%  & 3.50\%   \\
    Vendor E (Chip E)  & GLM-5   & 81.00\%  & 210.00\% \\
    \bottomrule
  \end{tabular*}
  \vspace{2pt}
  
\end{table}

A more striking example of this hardware asymmetry is observed with Vendor E (Chip E) running the GLM5 model. On a per-device basis, its optimal Prefill throughput reaches approximately 81\% of the H100 baseline. However, for the memory-bound Decode phase, the same chip achieves an impressive 210\% of the H100's performance. This extreme contrast—a chip that is moderately competitive in Prefill but vastly superior in Decode—highlights the core inefficiency of homogeneous clusters. Confining such a chip to a coupled deployment wastes its exceptional Decode capabilities, while utilizing baseline hardware for Decode fails to exploit available architectural advantages.

\subsection{Non-Uniform Transfer Latency under agentic Workloads}
\label{subsec:non_uniform_latency}

Although DRC can theoretically reduce cross-datacenter bandwidth requirements by up to 10$\times$ under an average prefix hit rate of 90\%, the absolute KV transfer latency varies significantly depending on the underlying workload distribution. In our preliminary experiments, we explore how the unique characteristics of agentic workloads shape the latency distribution and resource utilization during the Ethernet-based KV transfer phase, demonstrating that extreme transfer latencies are highly isolated to specific request types rather than being a global phenomenon.

Real-world agentic workloads are characterized by a highly skewed, broad distribution of prefix hit rates. While the average hit rate might be 90\%, the distribution spans a wide range from over 99\% down to 2\%. Crucially, this distribution is heavily concentrated at the higher end, with approximately 70-80\% of requests exhibiting ultra-high hit rates (e.g., $>$95\%), accompanied by a small fraction of low-hit-rate requests. This is fundamentally different from a uniform distribution where all requests hover tightly around the 90\% hit rate. 

\begin{figure}[htbp]
    \centering
    \includegraphics[width=\textwidth]{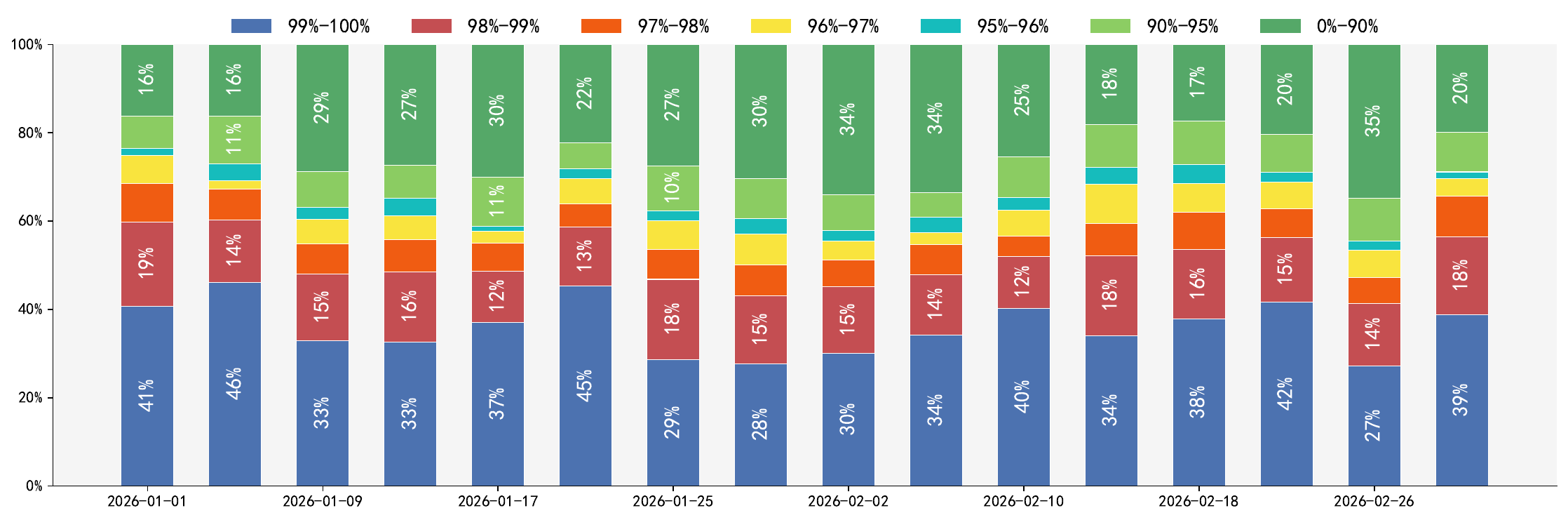}
    \caption{Distribution of requests across different hit rate intervals over time. The figure illustrates the request proportions for seven hit rate intervals ranging from 0\%-90\% up to 99\%-100\% during January to February 2026.}
    \label{fig:hit_rate_dist}
\end{figure}

To evaluate the impact of this highly skewed distribution, we conducted micro-benchmark tests for Ethernet-based KV cache transfer upon a 20Gbps cross-datacenter dedicated line under full-capacity conditions (Requests Per Second, RPS = 80). We compared two scenarios both targeting a 90\% average hit rate: (1) a \textit{Wide\&Skewed-Distribution} scenario reflecting real-world agentic workloads (hit rates ranging from 2\% to 99\% following a highly skewed Beta distribution), and (2) a \textit{Narrow\&Even-Distribution} scenario (hit rates ranging from 85\% to 95\% following a uniform distribution).

\begin{table}[htbp]
  \centering
  \caption{Impact of Hit Rate Distribution on Transfer Latency and System Saturation (RPS=80, Average Hit Rate = 90\%)}
  \label{tab:latency_distribution}
  \begin{tabular*}{\textwidth}{@{\extracolsep{\fill}}lcc@{}}
    \toprule
    \textbf{Metric} & \textbf{\makecell{Wide\&Skewed Distribution \\ (agentic Workload)}} & \textbf{\makecell{Narrow\&Even Distribution \\ (Uniform-like Workload)}} \\
    \midrule
    Hit Rate Range & 2\% $\sim$ 99\% & 85\% $\sim$ 95\% \\
    Per-req KV Payload Range & 0.008 GB $\sim$ 0.280 GB & 0.018 GB $\sim$ 0.047 GB \\
    \midrule
    Transfer Latency (P50) & \textcolor[HTML]{006400}{248 ms} & \textcolor[HTML]{8B0000}{1,210 ms} \\
    Transfer Latency (P99) & 18,883 ms & 25,542 ms \\
    Queue Latency (Avg) & \textcolor[HTML]{006400}{0.31 ms} & \textcolor[HTML]{8B0000}{1,682 ms} \\
    Connection Pool Utilization & \textcolor[HTML]{006400}{63.1\%} & \textcolor[HTML]{8B0000}{100.0\%} \\
    \midrule
    Dominant Performance Issue  & Tail-blockage on large requests & Connection pool exhaustion \\
    \bottomrule
  \end{tabular*}
  \vspace{2pt}
  
\end{table}

Our experiments (Table~\ref{tab:latency_distribution}) reveal two critical insights: First, \textbf{high transfer latencies are strictly isolated to low-hit-rate requests}. In the Wide\&Skewed scenario, the vast majority of high-hit-rate requests carry minimal payloads. These small packets exploit TCP fairness to bypass congestion, maintaining a remarkably low P50 latency ($\sim$248 ms). Conversely, the P99 latency spikes to $\sim$18.8s, which is exclusively attributable to the rare, massive payloads of low-hit-rate requests. Second, the skewed agentic distribution inherently \textbf{prevents TCP connection pool exhaustion}. The abundance of rapid-completing high-hit-rate requests continuously frees up TCP links, sustaining a high pool turnover rate ($\sim$63\% utilization). In contrast, the Narrow\&Even scenario lacks these lightweight requests to relieve pressure. Uniformly prolonged transfers cause concurrent connections to accumulate rapidly, fully exhausting the pool (100\% utilization). This triggers severe queuing (average 1,682 ms) and globally degrades both P50 and P99 latencies. Note that the reported transfer latencies exclude queue latency.

\subsection{Related Works}

PD disaggregation has been extensively studied to mitigate interference between the prefill and decode phases in LLM serving. Early works, such as DistServe \cite{Zhong2024DistServeDP} and Splitwise \cite{Patel2023SplitwiseEG}, primarily focus on intra-cluster disaggregation, leveraging high-speed RDMA networks to transfer KV cache. Recently, cross-datacenter inference has emerged as a promising paradigm to leverage geographically dispersed, heterogeneous hardware resources.  

Notably, PrfaaS (Prefill-as-a-Service) \cite{qin2026prefill}, introduced by MoonShot AI and Tsinghua KVCache.AI, serves as a highly representative cross-datacenter PD disaggregation architecture. While both PrfaaS and our PDD embrace the paradigm of geographically dispersed cross-datacenter disaggregation, they tackle fundamentally disparate challenges based on distinct technical assumptions. This orthogonality enables the seamless, synergistic integration of both approaches, facilitating the deployment of large-scale heterogeneous inference systems. Table~\ref{tab:pdd_vs_prfaas} delineates these key differences across several dimensions.

The divergence in fundamental motivation highlights the contrast between PrfaaS and PDD. PrfaaS mitigates SLA interference by specially designating remote instances for long-context prefilling, using a context-length-centric scheduler to balance cluster loads and prevent Ethernet overload. Its feasibility relies on Linear models, where per-token KV cache size decreases as context length grows, making remote transmission viable. Conversely, PDD harnesses the cost-effectiveness of heterogeneous hardware across datacenters to satisfy the strict SLAs of agentic workloads. Instead of relying on specific architectures, it capitalizes on the intrinsic, highly skewed KV cache hit-rates of agentic applications (averaging 90\%, with many exceeding 95\%). Through latency masking and multi-level pipeline orchestration, PDD overlaps cross-datacenter KV transfers exclusively for low-hit-rate outliers to hide transfer delays.

Ultimately, PDD and PrfaaS are orthogonal in their design philosophy: PrfaaS optimizes SLA fairness through request scheduling, whereas PDD optimizes end-to-end Service Level Objectives (SLO) for cross-datacenter heterogeneous inference via latency masking. Because they address different bottlenecks, they can be naturally integrated. A unified system could employ PrfaaS's scheduling logic to manage long/short request interference while deploying PDD's three-tier pipeline to ensure cross-datacenter inference remains performant and SLA-compliant.

\begin{table}[htbp]
\small
\centering
\caption{Comparison between PDD and PrfaaS across key dimensions}
\label{tab:pdd_vs_prfaas}
\renewcommand{\arraystretch}{1.3}
\begin{tabular}{p{0.12\textwidth} p{0.38\textwidth} p{0.42\textwidth}}
\toprule
\textbf{Dimension} & \textbf{PrfaaS} & \textbf{PDD} \\
\midrule
\textbf{Target\newline Problem} & 
To alleviate SLA interference between long and short requests in the prefill phase. & 
For high Benefit-Cost Ratio via Heterogeneous Cross-Datacenter under SLA Guarantees. \\
\midrule
\textbf{Core\newline Techniques} & 
Request scheduling and routing based on context length. & 
Latency masking and multi-level pipeline tuning to hide cross-datacenter transfer delays. \\
\midrule
\textbf{Key\newline Dependency} & 
Linear models, where per-token KV cache size decreases as context length grows. & 
agentic workload characteristics: high \& non-uniform KV cache hit rates (avg. 90\%). \\
\midrule
\textbf{Primary\newline Goal} & 
Prevent long requests from monopolizing resources, thereby improving overall service level. & 
Mitigate wide-area Ethernet bottlenecks to achieve SLA-compliance on cost-effective heterogeneous infra. \\
\midrule
\textbf{Relationship} & 
\multicolumn{2}{c}{Orthogonal and complementary; can be integrated seamlessly.} \\
\bottomrule
\end{tabular}
\end{table}

%% file: design.tex
\section{Design}
\label{sec:design}

\subsection{PDD Architecture Overview}
\label{subsec:overview}

\begin{figure}[htbp]
    \centering
    \includegraphics[width=\textwidth]{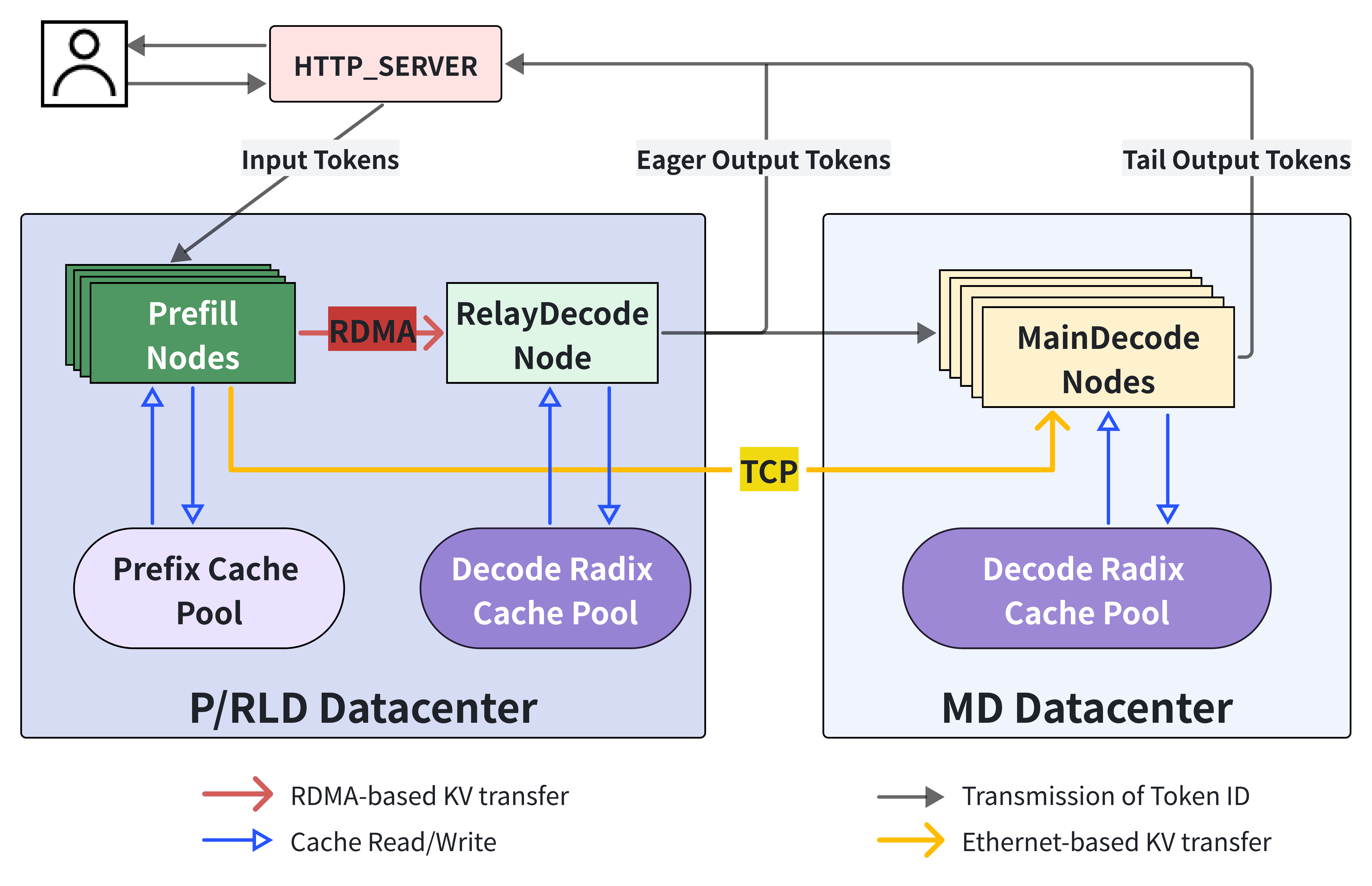}
    \caption{Overview of the PDD three-tier cross-datacenter PD disaggregation architecture. P and RLD instances are co-located within the same datacenter connected via high-speed RDMA, while MD instances are remotely connected via wide-area Ethernet.}
    \label{fig:pdd_architecture}
\end{figure}

As illustrated in Figure~\ref{fig:pdd_architecture}, PDD disaggregates the LLM inference lifecycle into three distinct instances distributed across geographically dispersed datacenters:
\begin{itemize}
    \item \textbf{Prefill (P) Instance}: Responsible for processing the prompt and generating the initial KV cache. P instances are deployed in a primary datacenter equipped with compute-heavy hardware.
    \item \textbf{RelayDecode (RLD) Instance}: Co-located with the P instance in the same datacenter and interconnected via high-speed RDMA networks. The RLD instance is specifically designed to eagerly generate decoding output tokens to mask cross-datacenter transfer latency.
    \item \textbf{MainDecode (MD) Instance}: Deployed in a geographically distant datacenter, connected to the primary datacenter via low-cost wide-area Ethernet (typically $<100$ Gbps). The MD instance handles the remaining bulk of the token generation phase.
\end{itemize}

The basic workflow initiates when the P instance receives a request and executes the prefill phase to generate the KV cache. If the cache hit-rate falls below a threshold $H$, the P instance concurrently dispatches the KV cache across two paths: to the RLD instance via high-bandwidth low-latency RDMA (taking tens of milliseconds) and to the MD instance via TCP over wide-area Ethernet (taking seconds). Leveraging the rapid RDMA transfer, the RLD instance immediately commences decoding, streaming the initial results—denoted as \texttt{Eager Output Tokens} in Figure~\ref{fig:pdd_architecture}—to the user to minimize perceived latency. Finally, upon completing the cross-datacenter KV transfer via much slower Ethernet, the MD instance seamlessly assumes the decoding process, ensuring uninterrupted token generation for the remainder of the request.

\subsection{Core Mechanism: Latency Masking via Extend-Decode Handoff}
\label{subsec:latency_masking}

While the PDD architecture leverages the RLD instance to eagerly generate initial tokens for transfer latency masking, a critical challenge arises once the MD instance completes KV cache reception: \texttt{how to seamlessly hand off the decoding task from RLD to MD}. 

To avoid the network overhead and H2D/D2H copies associated with transmitting the incremental KV cache from RLD to MD, the RLD instance only sends the CPU-resident token IDs to the MD instance. Upon receiving these tokens, the MD instance recomputes the incremental KV cache from these tokens while simultaneously generating the next output token—a process we term \textbf{Extend-Decode Handoff}. Specifically, Extend-Decode is a decoding paradigm commonly utilized in techniques such as Multi-Token Prediction (MTP) and Speculative Decoding. Unlike a standard decode step that takes a single token ID as input to produce the next token, it processes multiple input token IDs simultaneously to fill their corresponding KV cache and generate the subsequent output token. Due to the memory-bound nature of the decoding phase, Extend-Decode incurs marginal extra latency compared to a normal decode step; it adds little memory overhead, despite its significantly larger computational overhead. Furthermore, it is possible to fully overlap the extra latency introduced by Extend-Decode at the expense of heavier RLD load. To balance user experience against RLD load, PDD introduces two distinct Extend-Decode Handoff modes, as visually compared in Figure~\ref{fig:latency_masking_timeline}:

\begin{figure}[htbp]
    \centering
    \includegraphics[width=\textwidth]{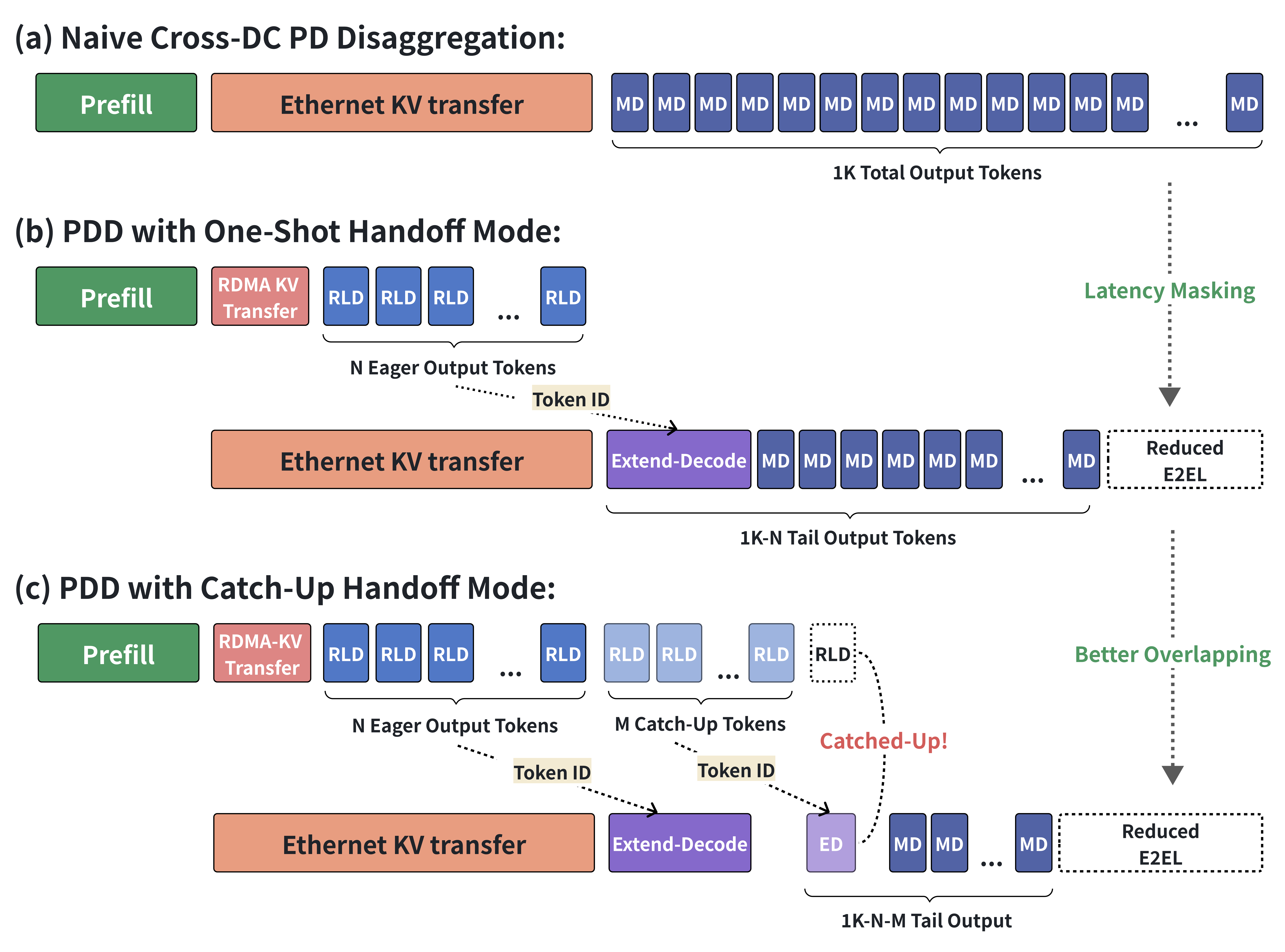}
    \caption{Timeline comparison of traditional cross-datacenter PD disaggregation versus PDD's latency masking mechanism. The figure illustrates PDD's two handoff modes: Catch-Up Handoff (gradual Extend-Decode with overlapped latency) and One-Shot Handoff (immediate control transfer). }
    \label{fig:latency_masking_timeline}
\end{figure}

\textbf{Catch-Up Handoff Mode.} In this mode, the MD instance performs Extend-Decode step-by-step, gradually catching up to the RLD's progress. Once MD's output progress catches-up with RLD's latest output, MD would seamlessly take over the token generation and replace RLD. The primary advantage of this mode is that it is completely transparent to the user; there are no observable fluctuations in OTPS (Output Tokens Per Second) and no temporary output stalls. However, to mask the computational time required for the MD's Extend-Decode Handoff phase, the RLD instance must continue generating a supplementary set of \textit{Catch-Up Tokens} before relinquishing control. The existence of these Catch-Up Tokens slightly increases the compute and memory load on the RLD instance.

\textbf{One-Shot Handoff Mode.} Alternatively, the RLD instance can transfer the entirety of its generated eager output tokens to the MD instance in a one-shot style and immediately evict the request, ceasing all further computation. The MD instance must then independently execute the Extend-Decode computation without latency overlapping. While this mode significantly reduces the load of RLD instance by minimizing its active inference duration, it introduces a trade-off: if the Extend-Decode overhead is substantial, the user may experience a temporary "frozen output" phenomenon, which also negatively impacts the overall OTPS.

\textbf{Overhead Analysis.} Both handoff modes replace the bulky incremental KV cache transmission with negligible Token ID transfers. The local recomputation on the memory-bound MD instance incurs minimal overhead, which is far more efficient and reliable than Ethernet-based KV transmission. PDD can dynamically alternate between these two strategies based on the real-time load of the RLD instance. For example, when RLD utilization is moderate and spare compute capacity is available, the system defaults to Catch-Up Handoff to guarantee a seamless, high-quality user experience.

\subsection{Multi-Stage Pipeline Orchestration}
\label{subsec:orchestration}

In PDD, the introduction of the RLD tier creates new possibilities and challenges for pipeline orchestration. In a conventional PD disaggregated architecture, the workflow is deterministic: a request must sequentially pass through the Prefill, KV cache transfer, and Decode phases. However, the PDD architecture introduces flexibility, allowing three distinct pipeline orchestrations capable of serving the complete lifecycle of any request: \textbf{P-RLD}, \textbf{P-MD}, and \textbf{P-RLD-MD}. As illustrated in Figure~\ref{fig:pdd_architecture}, the P-RLD-MD pipeline is inherently more complex, encompassing multiple stages including P, RDMA transfer, TCP transfer, RLD, and MD. In contrast, the P-RLD and P-MD pipelines are simpler, each comprising only three sequential stages.

\begin{figure}[htbp]
    \centering
    \includegraphics[width=\textwidth]{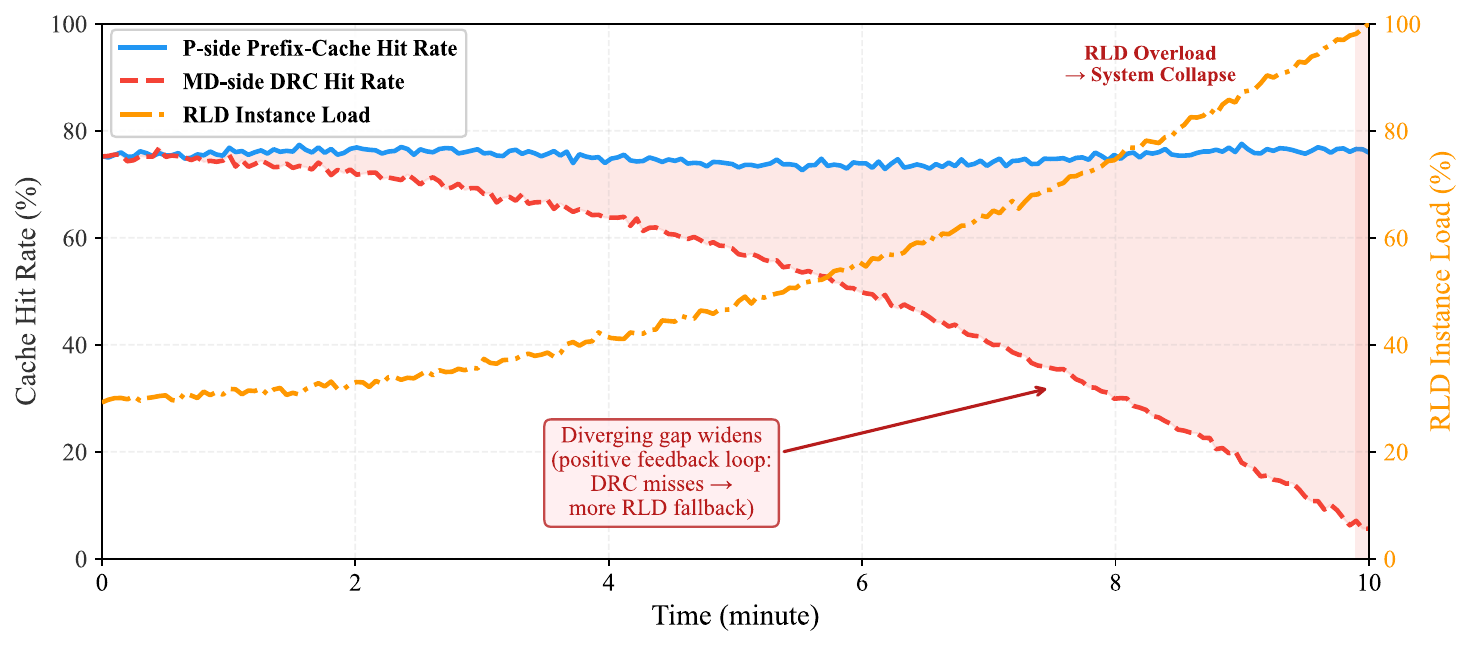}
    \caption{Impact of routing low-hit-rate requests (e.g., $<50\%$) through the P-RLD pipeline exclusively. Terminating KV cache transfers at RLD prevents the MD-side DRC from updating, causing its hit rate to progressively fall behind the P-side prefix-cache hit rate. This discrepancy eventually backfires, shifting disproportionate load back onto the RLD instances and risking system collapse.}
    \label{fig:decreasing_hit_rate}
\end{figure}

To effectively orchestrate these stages in real-world Agentic MaaS scenarios, we first establish a core principle: in the PDD cross-cluster PD disaggregated inference architecture, the RLD tier exists \emph{solely} to mask inter-cluster transfer latency and reduce TTFT. Accordingly, we favor \textbf{provisioning RLD with the minimum feasible hardware resources}. This minimal-resource posture, in turn, imposes stringent demands on how requests occupy and interact with RLD instances, leading to two subsequent criteria:
\begin{itemize}
    \item \textbf{Time-Bounded Occupation:} Given the unpredictable nature of LLM output lengths, requests with exceptionally long outputs must not indefinitely occupy the compute \& memory resources of RLD instances.
    \item \textbf{Cache Consistency:} As depicted in Figure~\ref{fig:decreasing_hit_rate}, the KV cache generated by P should ideally be transferred to the MD instance rather than terminating at the RLD instance.
\end{itemize}

Guided by the core principle and the two criteria above, the PDD architecture adopts only two pipeline orchestrations—\textbf{P-MD} and \textbf{P-RLD-MD}—explicitly excluding the P-RLD pipeline. Drawing upon the workload characteristics analyzed in Section~\ref{subsec:non_uniform_latency}, over $70\%$ requests are routed directly through the P-MD pipeline without any RLD intervention. The remaining $20\%$ to $30\%$ of low-hit-rate requests are processed via the P-RLD-MD pipeline. Because the MD instance ultimately assumes the decoding task, the maximum output length handled by the RLD instance is strictly bounded and can be dynamically adjusted based on real-time load conditions. This design inherently prevents long-output requests from monopolizing RLD compute and storage resources. Furthermore, since both active pipelines conclude at the MD stage, the architecture guarantees parity between the MD-side DRC hit rate and the P-side prefix-cache hit rate.

\subsection{Implications for Inference Infrastructure}
\label{subsec:infra_vision}

Beyond optimizing pipeline orchestration, the PDD architecture provides novel perspectives for the construction of future inference infrastructure. Among these, the most significant and insightful shift is that PDD enables low-cost, fine-grained heterogeneous deployment within a single datacenter.

\textbf{Low-Cost, Fine-Grained Heterogeneous Deployment.}
In a traditional intra-DC heterogeneous PD disaggregation system, accommodating both compute-bound prefill and memory-bound decode typically requires building massive, monolithic heterogeneous clusters. This approach introduces large amounts of secondary hardware (e.g., memory-access-efficient chips) alongside primary compute accelerators within a single RDMA domain. Such large-scale heterogeneity not only incurs extremely high cost but also suffers from severe inflexibility; when workload patterns shift, the fixed, rigid ratio of heterogeneous machines could easily lead to substantial resource waste.

The PDD architecture fundamentally resolves this bottleneck. Because the RLD instance only processes a small fraction of initial decoding tokens and merely works for low-hit-rate requests, it requires only moderate memory bandwidth and very little FLOPs. Consequently, providers can deploy a minimal fraction of memory-access-efficient chips for RLD instances, while allocating the vast majority of resources to compute-heavy P instances. The bulk of the decoding workload is then seamlessly offloaded to MD instances in a remote datacenter. 

\begin{figure}[htbp]
    \centering
    \begin{subfigure}[b]{0.375\textwidth}
        \centering
        \includegraphics[width=\textwidth]{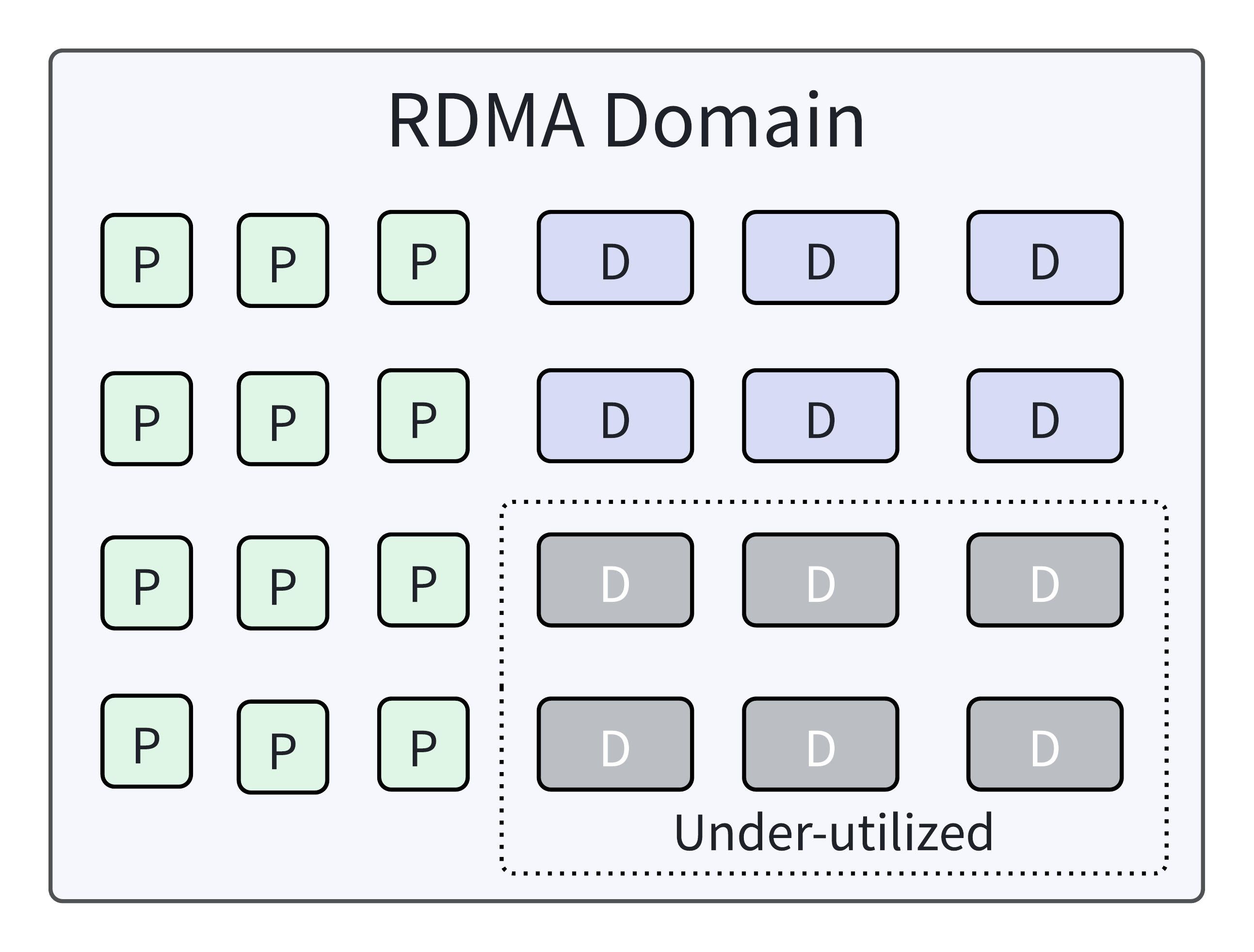}
        \caption{Traditional Intra-DC Heterogeneous PD}
        \label{fig:pd_left}
    \end{subfigure}
    \hfill
    \begin{subfigure}[b]{0.585\textwidth}
        \centering
        \includegraphics[width=\textwidth]{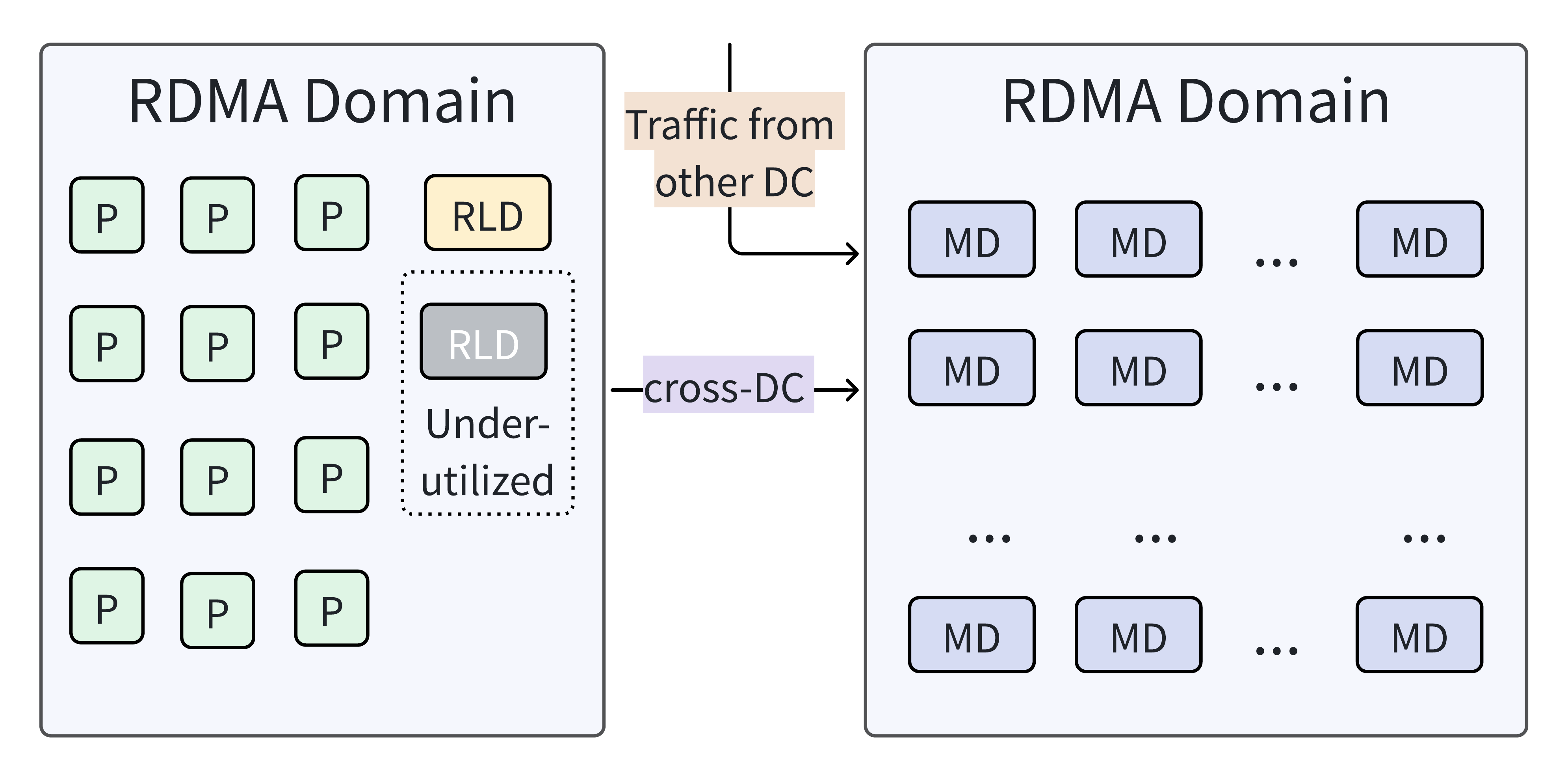}
        \caption{PDD Fine-Grained Cross-DC Heterogeneous Deployment}
        \label{fig:pdd_right}
    \end{subfigure}
    \caption{Comparison between traditional large-scale intra-DC heterogeneous PD deployment and the PDD fine-grained heterogeneous strategy. PDD minimizes heterogeneous cost and mitigates resource waste under shifting workloads.}
    \label{fig:hetero_compare}
\end{figure}

As illustrated in Figure~\ref{fig:hetero_compare}, this ultra-small-scale heterogeneous approach offers two major advantages:
\begin{itemize}
    \item \textit{Minimal cost:} Introducing only a tiny fraction of diverse accelerators drastically reduces capital expenditure compared to traditional intra-DC large-scale heterogeneous clusters.
    \item \textit{Higher Utilization:} The small number of RLD chips mitigates the resource waste associated with traffic pattern shift and hardware evolution.
\end{itemize}

\textbf{Flexible Cross-DC Topologies.}
Additionally, the decoupling of P/RLD and MD naturally extends to highly flexible cross-DC topologies. Based on the scalability of wide-area Ethernet, PDD enables a vast array of interconnection modes between P/RLD and MD clusters. Geographically dispersed primary datacenters (hosting P and fine-grained RLD) can be interconnected to centralized, large-scale MD clusters, or they can form distributed many-to-many topologies tailored to specific traffic distributions. This versatility allows MaaS providers to horizontally scale compute-bound prefill capabilities locally while flexibly centralizing or distributing memory-bound decode workloads, maximizing global resource utilization without being constrained by a single datacenter's physical limits. 

%% file: evaluation.tex
\section{Evaluation}
\label{sec:evaluation}
This section presents preliminary experimental results to validate the feasibility and core advantages of the PDD architecture. It should be noted that these results are non-exhaustive and primarily serve to demonstrate the effectiveness of PDD's latency overlaping and cost-efficient heterogeneous deployment under Agentic workloads. Comprehensive system-level optimizations, further adaptation to other heterogeneous chips, and large-scale empirical evaluations are planned for future work.

\subsection{Experimental Setup}
\label{subsec:eval_setup}
\textbf{Workloads.} 
We evaluated PDD using \texttt{real-world Agentic workload traces} (we only utilize the overall statistics such as input/output length distribution and cache hit rates, no user-sensitive information is involved) to reflect the workload characteristics discussed in Section~\ref{subsec:non_uniform_latency}. The Agentic workload features an average prefix cache hit rate of 90\% with a highly skewed Beta distribution (ranging from 2\% to 99\%), where over 70\% of requests have ultra-high hit rates. Output lengths are typically short (30\% of requests produce fewer than 100 tokens), with an average output length of 1000 tokens. 

\textbf{Models and Hardware.} 
We conducted tests on DeepSeek-V4-pro. For the PDD deployment, the primary datacenter (hosting P and RLD instances) is equipped with $43$ NVIDIA H100 nodes (344 GPUs) compute-heavy cluster for 10 P instances and a single RLD instance. The remote datacenter hosts the MD instances, deployed on a cluster with $32$ NVIDIA H200 nodes (256 GPUs). The two datacenters are interconnected via $2\times$10 Gbps wide-area Ethernet dedicated line, simulating cost-effective cross-datacenter connections. Intra-DC communication between P instances and the RLD instance utilizes a standard 400 Gbps RDMA network. For the intra-DC PD Disaggregation deployment, we use a cluster encompassing $88$ H100 nodes (704 GPUs) for both P and D instances.

\textbf{Baselines and Metrics.} 
To evaluate the effectiveness of latency overlaping, we compare PDD against a traditional Cross-Datacenter PD Disaggregation baseline (denoted as \textit{CDC-PD}). In this setup, DRC is enabled properly but RLD latency overlaping mechanism is not employed. To evaluate the cost-efficiency of PDD, we compare PDD against intra-DC PD Disaggregation (denoted as \textit{IDC-PD}). Key metrics monitored in the evaluation include TTFT, TPOT, and RPS. 

\subsection{End-to-End Performance and Latency Overlap}
\label{subsec:eval_e2e}
\begin{figure}[htbp]
  \centering
  \begin{subfigure}{0.415\columnwidth}
    \centering
    \includegraphics[width=\linewidth]{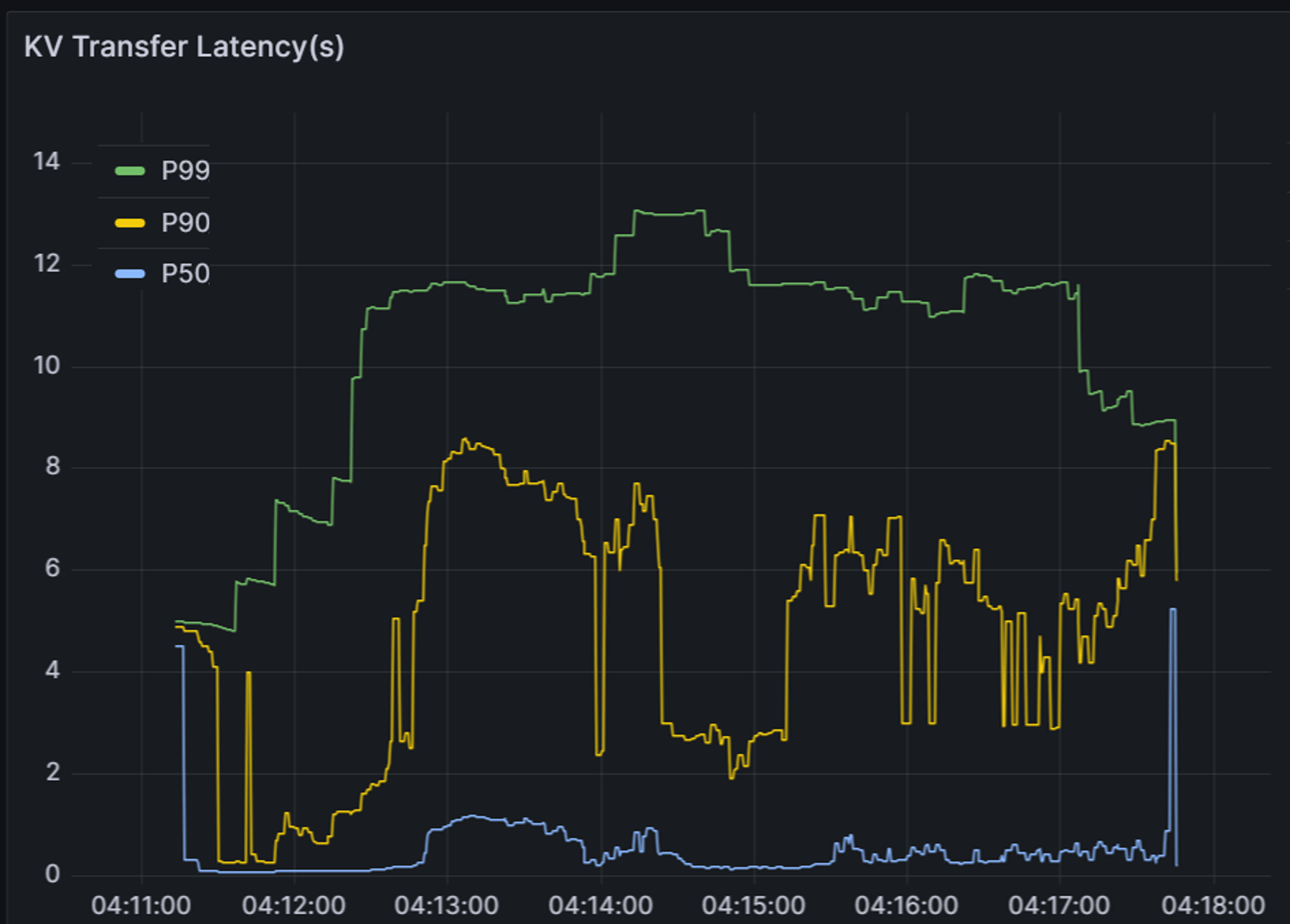}
    \caption{Transfer latency variation of CDC-PD.}
    \label{fig:transfer_latency}
  \end{subfigure}
  \hfill
  \begin{subfigure}{0.545\columnwidth}
    \centering
    \includegraphics[width=\linewidth]{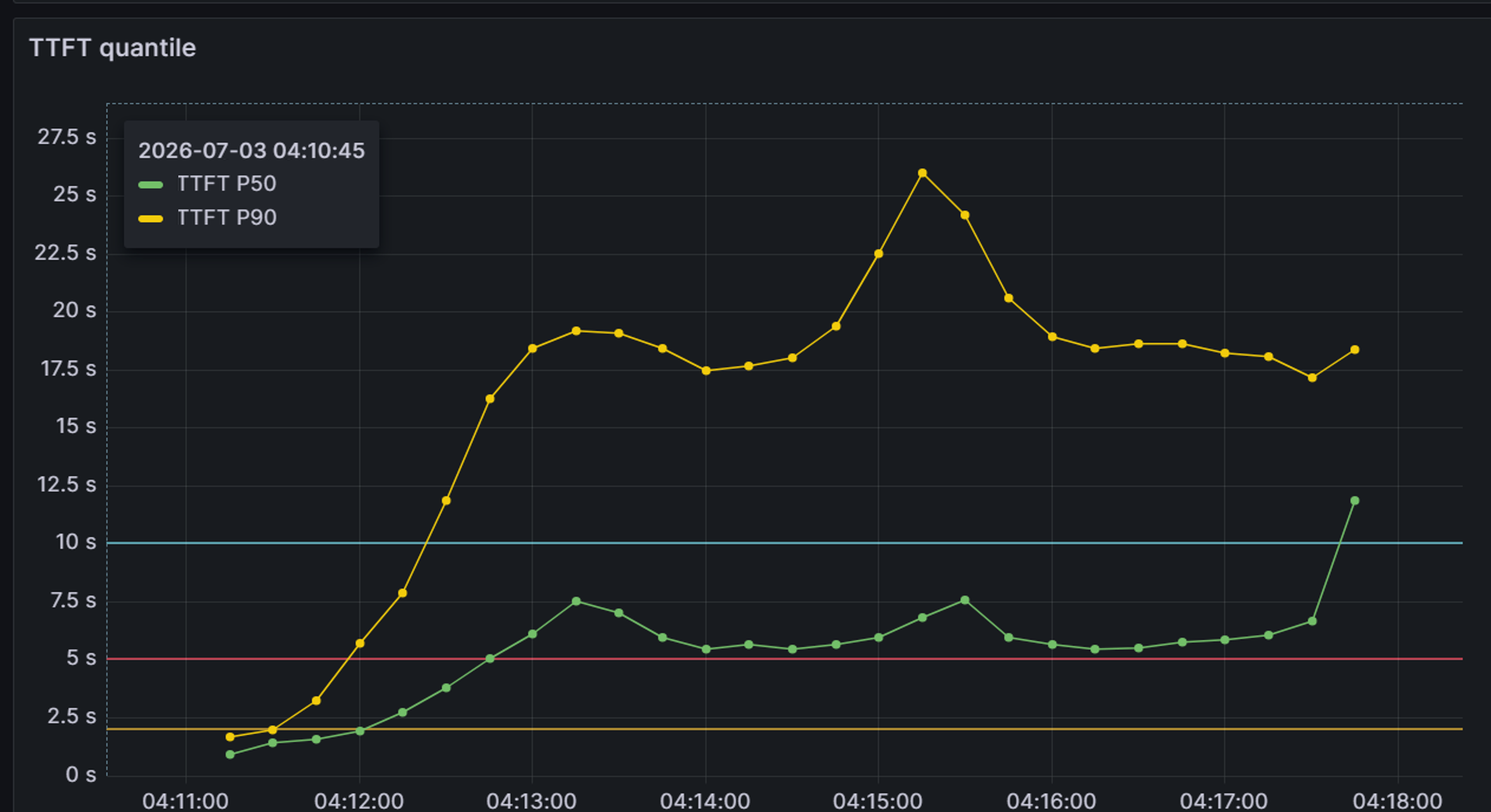}
    \caption{TTFT variation of CDC-PD.}
    \label{fig:ttft}
  \end{subfigure}
  \caption{Latency characteristics of the CDC-PD baseline under Agentic workloads.}
  \label{fig:cdc_pd_latency}
\end{figure}

We first evaluate the effectiveness of PDD's latency overlaping mechanism in reducing the user-perceived transfer latency (reflected in TTFT) under Agentic workloads. As shown in Table~\ref{tab:ttft_comparison} and Figure~\ref{fig:cdc_pd_latency}, the naive CDC-PD baseline suffers from severe TTFT degradation, with P90 latency spiking to over 18 seconds due to the Ethernet-based KV cache transfer bottleneck. Figure~\ref{fig:transfer_latency} reveals that CDC-PD's transfer latency fluctuates dramatically across requests, with frequent multi-second spikes caused by unpredictable Ethernet congestion; these spikes directly propagate to the user-perceived TTFT shown in Figure~\ref{fig:ttft}, where the latency tail stretches well beyond 20 seconds. In contrast, PDD successfully overlaps this transmission delay, exhibiting much better P99 and P90 TTFT at the expense of an additional RLD instance.

Because the RLD instance immediately initiates decoding upon receiving the KV cache via high-speed RDMA, PDD maintains a stable and low TTFT comparable to intra-DC homogeneous deployments. As evidenced by the high-variance transfer latency in Figure~\ref{fig:transfer_latency}, CDC-PD's Ethernet-based transmission is the root cause of the TTFT tail inflation observed in Figure~\ref{fig:ttft}; PDD sidesteps this bottleneck entirely. For the vast majority of high-hit-rate requests (which terminate via the P-MD pipeline with minimal KV payloads), the transfer latency is negligible. For low-hit-rate outliers, the RLD instance actively generated Eager Output Tokens while the slow Ethernet transfer occurs in the background, ensuring that the user perceives continuous token generation without multi-second stalls. Overall, PDD reduces the P90 TTFT by approximately 46\% compared to the CDC-PD, effectively mitigating the cross-datacenter latency bottleneck.

\begin{table}[htbp]
  \centering
  \caption{TTFT Comparison between PDD and CDC-PD}
  \label{tab:ttft_comparison}
  \begin{tabularx}{0.7\linewidth}{l*{5}{>{\centering\arraybackslash}X}}
    \toprule
    Method & P99 & P90 & P50 & Median & Mean \\
    \midrule
    CDC-PD & 29.7s & 18.3s & 6.7s & 6.7s & 9.2s \\
    PDD    & 14.4s & 9.8s  & 4.2s & 4.2s & 5.1s \\
    \bottomrule
  \end{tabularx}
\end{table}

\subsection{Microbenchmark: RLD Load Analysis and Handoff Overhead}
\label{subsec:eval_handoff}

\begin{figure}[htbp]
    \centering
    \includegraphics[width=\textwidth]{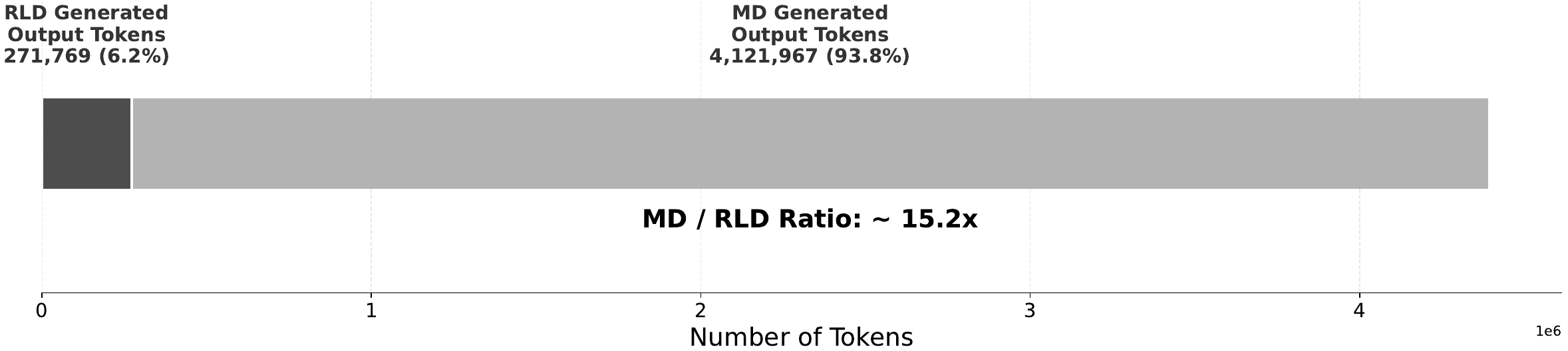}
    \caption{Comparison of output token load distribution between RLD and MD. The horizontal bar chart illustrates a notable load imbalance, where MD dominates the workload by generating 4,121,967 tokens (93.8\% of the total), whereas RLD only outputs 271,769 tokens (6.2\%), highlighting a ~15.2x disparity in processing load.}
    \label{fig:RLD_ratio}
\end{figure}

To evaluate the load status of RLD, we examine the percentage of output tokens generated by RLD against total output tokens under the scenario of 64K input (90\% hit rate) and 1K output. As illustrated in Figure~\ref{fig:RLD_ratio}, the load of RLD is far less than the load of MD instances.

Furthermore, we compared the two handoff modes:
\begin{itemize}
    \item \textbf{Catch-Up Handoff Mode} achieved a seamless OTPS curve with zero observable output stalls. The RLD instance maintained active inference for slightly longer to generate Catch-Up Tokens, increasing its utilization (reflected by RLD's average output length) by roughly 10\% compared to One-Shot mode.
    \item \textbf{One-Shot Handoff Mode} minimized RLD resource occupation, freeing up memory and compute capacity faster. However, it introduced a brief, temporary OTPS dip (lasting $<0.5$ seconds for 1K-token outputs) during the MD's isolated Extend-Decode phase. 
\end{itemize}

To evaluate the overhead of the Extend-Decode Handoff mechanism, we conducted micro-benchmarks to compare the naive approach of transmitting incremental KV caches over Ethernet against our proposed local recomputation approach (transmitting lightweight Token IDs followed by MD-side recomputation). 

The results confirm that replacing bulky KV cache transmission with local recomputation is highly efficient and robust. As shown in Table~\ref{tab:kv_vs_recomp}, transmitting KV caches over Ethernet suffers from severe latency fluctuations. In our tests for a 100-token handoff, the average latency was 185.4ms, but peaked at 512.6ms due to network fluctuations and connection pool queueing, with a high standard deviation of 96.3ms. Furthermore, this method incurs an additional 24.5ms overhead for H2D/D2H memory copies. 

In contrast, our local recomputation approach transmits only a few kilobytes of Token IDs (e.g., 1.5 KB). This transmission is fully asynchronous—RLD sends each token ID to MD as it is generated—making its network overhead negligible. The local Extend-Decode recomputation for 100 tokens takes an average of 305.2ms. Although the baseline average time is slightly higher than a clear-network KV transmission, its maximum latency is capped at 321.4ms with a minimal standard deviation of 4.8ms. Crucially, it is immune to network fluctuation and queueing interference and eliminates extra H2D/D2H overheads entirely.

\begin{table}[htbp]
    \centering
    \caption{Microbenchmark: Handoff Overhead Comparison for a 100-Token Sequence}
    \label{tab:kv_vs_recomp}
    \begin{tabularx}{0.8\linewidth}{l>{\centering\arraybackslash}X>{\centering\arraybackslash}X}
        \toprule
        \textbf{Metric} & \textbf{KV Cache Trans.} & \textbf{Local Recomp.} \\
        \midrule
        Payload Size (KB) & 4649 & 1.5 \\
        Min Latency (ms)  & 75.2 & 289.5 \\
        Avg. Latency (ms) & 185.4 & 305.2 \\
        Max Latency (ms) & 512.6 & 321.4 \\
        Latency Std. Dev. (ms) & 96.3 & 4.8 \\
        H2D/D2H Overhead (ms) & 24.5 & --- \\
        \bottomrule
    \end{tabularx}
\end{table}

\subsection{Cost-Efficiency and Benefit-Cost Ratio of Heterogeneity}
\label{subsec:eval_cost}
Finally, we evaluate the cost-efficiency of PDD’s cross-datacenter heterogeneous deployment compared to the traditional intra-DC homogeneous PD disaggregated cluster (IDC-PD). 

In our PDD deployment, compute-intensive H100 GPUs in the primary datacenter serve the P and RLD instances, where each P instance consumes $4$ H100 nodes and the RLD instance consumes $3$ H100 nodes. Concurrently, memory-bandwidth-optimized H200 GPUs are deployed in the remote datacenter to handle MD instances, each utilizing $2$ H200 nodes. In IDC-PD's deployment, each P instance consumes $4$ H100 nodes and each D instance consumes $3$ H100 nodes.

We adopt the Benefit-Cost Ratio (BCR)—defined as the ratio of SLA-compliant effective request throughput (goodput) to normalized hardware cost—as our primary metric, employing an SLA of 10s P90 TTFT and 50ms P90 TPOT (the actual latencies achieved is shown in Table~\ref{tab:sla_compliance}). Based on the resource unit prices listed in Table~\ref{tab:resource_pricing}, experiment results demonstrate that PDD achieves a \textbf{BCR up to 37.5\% higher} than the IDC-PD baseline (Table~\ref{tab:bcr_comparison}). This improvement is attributed to PDD’s capability to map each inference phase to the most suitable hardware across datacenters: the H100 cluster manages the compute-bound prefill and latency-overlaping RLD, while the remote H200 cluster absorbs the memory-bound massive decode workload.

\begin{table}[htbp]
\centering
\small

\begin{minipage}[t]{0.56\textwidth}
    \centering
    \caption{Resource Unit Price Description}
    \label{tab:resource_pricing}
    \begin{tabularx}{\linewidth}{@{}llX@{}}
        \toprule
        \textbf{Resource} & \textbf{Unit Price} & \textbf{Description} \\
        \midrule
        H100 GPU    & $X$              & $X$ is the current price of an $8\times$H100 \\
        H200 GPU    & $1.2\sim1.3X$ & H200 premium over H100 is $20\%\sim30\%$ \\
        \midrule
        \makecell[l]{10\,Gbps \\ Ethernet} & $0.35X$ & Cost for one cross-datacenter dedicated line \\
        \bottomrule
    \end{tabularx}
\end{minipage}%
\hfill
\begin{minipage}[t]{0.42\textwidth}
    \centering
    \caption{Latency Metrics}
    \label{tab:sla_compliance}
    \begin{tabular*}{\linewidth}{@{\extracolsep{\fill}}llccc@{}}
        \toprule
        \textbf{Metric} & \textbf{Method} & \textbf{P50} & \textbf{P90} & \textbf{P99} \\
        \midrule
        \multirow{2}{*}{TTFT} & IDC-PD & 3.9s    & 10.4s   & 17.2s  \\
                             & PDD          & 4.2s    & 9.8s    & 14.4s  \\
        \midrule
        \multirow{2}{*}{TPOT} & IDC-PD & 49.5ms  & 51.7ms  & 53.1ms      \\
                             & PDD          & 45.4ms & 47.9ms  & 49.7ms      \\
        \bottomrule
    \end{tabular*}
\end{minipage}

\end{table}

\begin{table}[htbp]
  \centering
  \caption{BCR Comparison between IDC-PD and PDD  (64K Context, 90\% Hit Rate, SLA-Compliant Goodput)}
  \label{tab:bcr_comparison}
  \begin{tabular*}{\textwidth}{@{\extracolsep{\fill}}lccc@{}}
    \toprule
    \textbf{Metric} & \textbf{\makecell{IDC-PD \\ (10P-16D)}} & \textbf{\makecell{PDD \\ (10P-1RLD-16MD)}} & \textbf{\makecell{Increment \\ PDD Over IDC-PD}} \\
    \midrule
    \#H100        & 88                         & 43                               & -45 \\
    \#H200        & ---                       & 32                               & +32 \\
    \midrule
    Ethernet Cost   & ---                         & $0.35X$                              & $+0.35X$ \\
    GPU Cost   & $88X$                       & $81.4X \sim 84.6X$                   & $-6.6X \sim -3.4X$ ($-7.5\% \sim -3.9\%$) \\
    Total Cost     & $88X$                       & $81.75X \sim 84.95X$                 & $-6.25X \sim -3.05X$ ($-7.1\% \sim -3.5\%$) \\
    \midrule
    RPS        & $9.0$                       & $11.5$                               & $+2.5$ ($+27.8\%$) \\
    BCR (RPS/\$) & $0.1023$                  & $0.1407 \sim 0.1354$                 & $+0.0384 \sim +0.0331$ ($+37.5\% \sim +32.4\%$) \\
    \bottomrule
  \end{tabular*}
  \vspace{2pt}
  
\end{table}

It is worth noting that these preliminary results do not fully capture the ultimate potential of PDD-enabled, cross-datacenter heterogeneous disaggregated inference. First, due to resource constraints, our testing was limited to a medium scale, utilizing 3 H100 nodes as the RLD instance—the minimal setup required for H100 GPUs to serve the Deepseek-V4-pro model. Given the relatively low RLD load at this scale, we expect its associated cost overhead to be better amortized in larger deployments, potentially yielding an even higher BCR for PDD. Second, our experiments only evaluated heterogeneity between H100 and H200 GPUs. While their cost-performance divergence across the P and D phases of Deepseek-V4-pro is significant, this is not the optimal pairing; integrating other specialized AI chips with H100 or H200 could unlock substantially greater BCR improvements. Finally, MTP was disabled and some other system-level optimizations for both phases were not fully applied (due to DRC related compatibility issues), resulting in a performance gap compared to State-of-the-Art (SOTA) implementations. Nevertheless, this lack of tuning does not undermine our core conclusion: PDD-style heterogeneous deployment is able to achieve higher BCR in SLA-compliant goodput compared to traditional homogeneous PD disaggregation.

%% file: discussion.tex
\section{Discussion}
\label{sec:discussion}

\textbf{Effectiveness Rooted in Workload Characteristics.} 
The superior performance of PDD is not merely a byproduct of advanced network optimization (which we did in extensively optimizing the TCP ransfer code in MoonCake Transfer Engine), but rather the result of deeply exploiting the intrinsic characteristics of Agentic workloads. The highly skewed distribution of prefix cache hit rates (averaging 90\% with over 70\% exceeding 95\%) fundamentally alters the network traffic profile. By routing the vast majority of high-hit-rate requests through the P-MD pipeline, the payload size is drastically minimized, circumventing TCP connection pool exhaustion and Ethernet congestion. Consequently, the system only needs to apply the latency masking mechanism to the isolated tail of low-hit-rate requests via the P-RLD-MD pipeline. By converting unreliable, bulky cross-datacenter KV cache transmissions into lightweight Token ID transfers followed by local recomputation, PDD transforms an unpredictable wide-area network bottleneck into a deterministic, compute-bound local operation.

\textbf{Limitations and Future Improvements.} 
While our preliminary evaluations demonstrate clear advantages, certain limitations exist that point to future research directions. First, due to hardware availability, our RLD deployment was constrained to a 3-node H100 setup. Given that RLD handles only 6.2\% of the total token generation load, we anticipate that in larger-scale deployments, the fixed RLD overhead will be better amortized, yielding even higher BCR. Second, system-level optimizations such as MTP were disabled in our current tests due to DRC compatibility constraints. Finally, our evaluation paired H100 with H200 GPUs. Exploring the integration of highly specialized, non-GPU AI accelerators (e.g., ASICs optimized for memory-bound decoding) holds the potential to unlock substantially greater cost-efficiency.

\textbf{Implications for Future MaaS Infrastructure.} 
Beyond algorithmic optimizations, PDD offers a paradigm shift for datacenter construction. Traditional intra-DC heterogeneous deployments require massive, rigid, monolithic clusters, incurring prohibitive cost and poor adaptability to shifting model architectures. PDD validates a "minimal local heterogeneity + flexible remote disaggregation" approach. By requiring only a tiny fraction of memory-access-efficient chips for RLD instances within a primary compute-heavy datacenter, and offloading the bulk of the decoding workload to geographically dispersed MD clusters via low-cost Ethernet, MaaS providers can achieve unprecedented elasticity. This fine-grained, cross-datacenter design establishes a scalable and economically viable blueprint for next-generation inference infrastructure.

%% file: conclusion.tex
\section{Conclusion}
\label{sec:conclusion}

The rapid evolution of LLMs and their diverse hardware requirements have made heterogeneous Prefill-Decode disaggregation a necessity for cost-efficient inference. However, deploying such systems across geographically dispersed datacenters introduces severe wide-area Ethernet bottlenecks, particularly for latency-sensitive Agentic workloads characterized by ultra-long contexts and high prefix hit rates. In this technical report, we proposed PDD, a novel three-tier cross-datacenter architecture designed to enable economical and flexible heterogeneous inference. PDD introduces a co-located RelayDecode instance that eagerly generates initial tokens via high-speed RDMA to mask the slow cross-datacenter KV transfer. To seamlessly hand off decoding to the remote MainDecode instance, we designed the Extend-Decode Handoff mechanism, which transmits only lightweight Token IDs and leverages local recomputation, effectively eliminating bulky and unreliable network transfers. Supported by a multi-stage pipeline orchestration that routes traffic based on cache hit rates, PDD prevents RLD overload while maintaining cache consistency.

Extensive evaluations on DeepSeek-V4-pro using real-world Agentic workload traces demonstrated the efficacy of our approach. Compared to the traditional Cross-Datacenter PD baseline, PDD successfully masked transmission latency, reducing P90 TTFT by approximately 46\% and capping maximum handoff latency at 321.4ms with minimal variance. Furthermore, compared to the Intra-DC homogeneous PD baseline, PDD's fine-grained heterogeneous mapping of compute-intensive and memory-bandwidth-optimized hardware achieved a Benefit-Cost Ratio up to 37.5\% higher in SLA-compliant goodput. Ultimately, PDD not only resolves the immediate transmission bottlenecks of cross-datacenter inference but also provides a highly scalable, adaptable, and cost-effective blueprint for future MaaS infrastructures.